\documentclass{pasj00}
\SetRunningHead{Tsujimoto \& Shigeyama}{Ba in SN1987A}

\def\cm3{{\rm ~cm}^{-3}}
\def\ltsima{$\; \buildrel < \over \sim\;$}
\def\ltsim{\lower.5ex\hbox{\ltsima}}
\def\gtsima{$\; \buildrel > \over\sim \;$}
\def\gtsim{\lower.5ex\hbox{\gtsima}}
\def\ms{$M_{\odot}$ }
\def\msp{$M_{\odot}$}

\author{Takuji \textsc{Tsujimoto}} 
\affil{National Astronomical Observatory, Mitaka-shi, 
Tokyo 181-8588, Japan}
\email{taku.tsujimoto@nao.ac.jp} 
\and
\author{Toshikazu \textsc{Shigeyama}}
\affil{Research Center for the Early Universe, Graduate 
School of Science, University of Tokyo, Bunkyo-ku, Tokyo 113-0033, 
Japan}
\email{shigeyama@astron.s.u-tokyo.ac.jp}

\title{SUPERNOVA SN 1987A REVISITED AS A MAJOR PRODUCTION SITE 
FOR $r$-PROCESS ELEMENTS}

\Received{10 August 2001}
\Accepted{19 September 2001}

\begin{document}
\maketitle

\begin{abstract}
The origin of nucleosynthesis products of rapid neutron capture
reactions (the $r$-process) is a longstanding astrophysical problem.
Recent analyses of elemental abundances for extremely metal-poor stars
shed light on the elemental abundances of individual supernovae.
Comparison of the abundance distributions of some extremely metal-poor
stars with those of the best-observed supernova SN 1987A clearly
indicates that the overabundances of barium and strontium found in SN
1987A that have been ascribed to the slow neutron capture process must
be results of $r$-process nucleosynthesis. The mass of freshly
synthesized barium in SN 1987A is estimated to be
$6\times10^{-6}\,M_\odot$ based on the observed surface abundance and
detailed hydrodynamical models for this supernova. These new findings
lead to the conclusion that 20 \ms stars, one of which is the progenitor
star of SN 1987A, are the predominant production sites for $r$-process
elements in the Galaxy and the $r$-process element donors for notable
neutron-capture-rich giant stars, CS22892-052 and CS31082-001.
\end{abstract}

\section{INTRODUCTION}

The existence of astrophysical $r$-process elements (elements
synthesized by rapid neutron capture process) in old metal-poor stars
has indicated that these elements are synthesized in the preceding
supernovae (SNe) originating in massive stars \citep{Truran_81}. The
elemental abundance determination of numerous Galactic halo stars with
extremely low metal abundances has opened the door to much more detailed
investigations of $r$-process production from SNe in the early
Galaxy \citep{McWilliam_95,Ryan_96}. However, which SNe produce
the $r$-process elements has not yet been agreed upon in spite of many
authors' attempts \citep{Mathews_92,Ishimaru_99,Tsujimoto_00,
Qian_01,Fields_01}.

Direct information about the production site for $r$-process elements
should come from abundance determination of the elements in the ejecta
of individual SNe. Absorption lines due to barium (Ba) and strontium
(Sr), $\sim$10\% of which are of $r$-process origin in the solar
elemental abundance, were detected in the spectra of SN 1987A
\citep{Williams_87}. A remarkable feature found by Mazzali, Lucy, \&
Butler (1992) was that the shape of the Ba absorption line indicated
lack of this element at the very surface of the ejecta. Thus the
observed Ba must have been synthesized inside the star and did not
exist in the interstellar matter (ISM) from which the star formed. The
abundances of Ba and Sr were derived by taking into account non-LTE
effects. The result showed that these elements are overabundant in
comparison with the abundance of the Large Magellanic Cloud
\citep{Hoeflich_88,Mazzali_92,Mazzali_95}.  The overabundances of Ba
and Sr in SN 1987A have led a few authors to argue that these elements
were synthesized by a slow neutron capture process (weak $s$-process)
operating in the core helium (He) burning stage of the progenitor star
(e.g., Prantzos, Arnould, \& Cass\'e 1988). However, the derived
abundance ratio of these two elements in SN 1987A was larger than the
corresponding solar ratio, i.e., Ba/Sr$\sim$2.5(Ba/Sr)$_\odot$. This
value is clearly inconsistent with the results of weak $s$-process
calculations \citep{Prantzos2_88} which predict a ratio significantly
smaller than the solar ratio such as 0.1(Ba/Sr)$_\odot
\leq$Ba/Sr$\leq$0.6(Ba/Sr)$_\odot$ (Fig.~1). Another fact casts doubt
on $s$-process origin: other SNe II-P such as SN 1985P, SN1990E, and
SN 1990H did not show any sign of the overabundance of Ba
\citep{Chalabaev_87,Mazzali_95}, though $s$-process calculations
predict the overabundance of Ba irrespective of the He core mass
(Prantzos et al.~1988). Together, these facts strongly suggest that
these elements in SN 1987A are not of $s$-process origin.

\begin{figure}
\begin{center}
\FigureFile(239bp,154bp){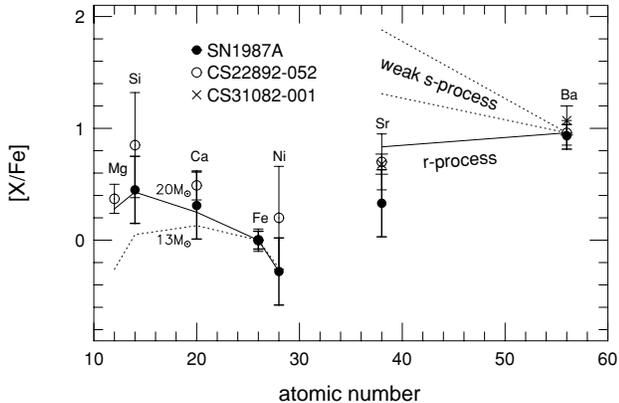}
\caption{Comparison of the elemental abundance distribution
in the ejecta of SN 1987A with that of the star CS22892-052. For Sr and
Ba, the abundance distribution of another neutron-rich giant star
CS31082-001 is also plotted together with the solid and dashed lines
showing the abundance distributions theoretically anticipated from the
$r$-process origin and $s$-process nucleosynthesis \citep{Prantzos_88},
respectively. These lines are drawn from the [Ba/Fe] corresponding to
the ratio of CS22892-052. For lighter elements between Mg and Ni, the
abundance curves predicted by a recent nucleosynthesis calculation for
20\ms (solid line) and 13\ms (dashed line) stellar models
\citep{Umeda_01} are also shown.}
\end{center}
\end{figure}

On the other hand, the derived Ba/Sr ratio can be reconciled with the
ratios of the same elements in two extremely metal-poor stars
CS22892-052 and CS31082-001 (Fig.~1). Since these elements in such
metal-poor stars must be results of $r$-process nucleosynthesis
\citep{McWilliam_98}, the derived ratio suggests that these elements
in SN 1987A might also be of $r$-process origin. If this is the case,
it is expected that the observed Ba (Sr) is synthesized during
explosion in the deepest layers of the ejecta where the matter is
exposed to an intense flux of neutrons. At the same place, a
radioactive element nickel 56 ($^{56}$Ni), which is one of the major
seed elements for $r$-process nucleosynthesis, is also
synthesized. The detection of $\gamma$-ray lines emitted from
$^{56}$Co at unexpectedly early times \citep{Matz_88} and
multi-dimensional hydrodynamical calculations (e.g., Fryxell, M\"uller,
\& Arnett 1991; Hachisu et al.~1990) suggest that the Rayleigh-Taylor
instabilities occurring at each composition interface in the
progenitor star after the passage of the blast wave have brought the
matter in the deep region near the mass-cut out to the surface regions
expanding at a few thousand km s$^{-1}$. This mixing process finishes
soon after the blast wave hits the stellar surface. The hydrodynamical
evolution would not change the relative abundances of these elements
inside each fluid element. Therefore it is likely that the spatial
distributions of these three elements are similar throughout the
explosion. With this in mind, we will discuss the amount of Ba
synthesized in SN 1987A with the support of detailed observations and
modelling for this supernova in the next section.

\section{SYNTHESIZED BARIUM MASS IN SN 1987A}

The progenitor star of SN 1987A was identified as a B3 I star
Sk-69$^\circ$202 \citep{Sonneborn_87}. Thus the initial mass is
estimated as $\sim$20 \ms and the radius as $\sim$40 $R_\odot$. To
estimate the stellar mass at explosion, one needs to consider the
evolutionary path to the B3 I star, because this type of star undergoes
mass loss due to the effects of stellar wind. Numerical calculations
\citep{Saio_88} suggest that the stellar mass at explosion is about 16
\msp.

\begin{figure}
\begin{center}
\FigureFile(262bp,274bp){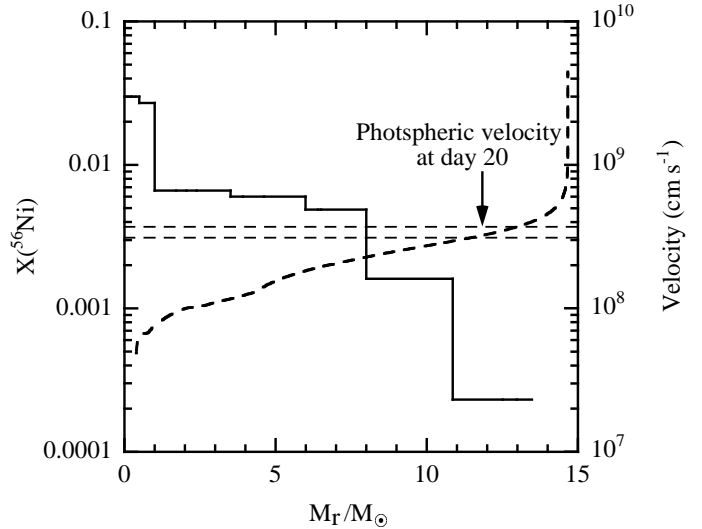}
\vspace{-2.5cm}
\caption{The solid curve displays the distribution of
$^{56}$Ni in SN 1987A inferred from X- and $\gamma$-ray observations as
a function of enclosed mass \citep{Kumagai_89,Shigeyama_90}. The dashed
curve displays the velocity distribution of a hydrodynamical model 14E1
of SN 1987A \citep{Shigeyama_90}. Two horizontal dashed lines show the
expansion velocities of the photosphere inferred from observations
\citep{Menzies_88} at day 20.}
\end{center}
\end{figure}

Multi-band photometric observations for SN 1987A have been used to
construct so-called bolometric light curves during the first 400 days
after explosion \citep{Menzies_88}. Comparison of hydrodynamical
models for this B3 I progenitor star explosion with the bolometric
light curve enables us to estimate the explosion energy, the amount of
freshly synthesized $^{56}$Ni, and the neutron star mass left behind
(e.g., Shigeyama \& Nomoto 1990). Furthermore, X- and $\gamma$-rays
from this supernova were detected. These emissions originate in
$^{56}$Ni decays and the subsequent $^{56}$Co decays. The spectra and
light curves of these emissions give us information about the
distribution of $^{56}$Ni. Monte-Carlo simulations of $\gamma$-ray
transfer in the ejecta derived the $^{56}$Ni distribution as a
function of the enclosed mass \citep{Kumagai_89} based on a
hydrodynamical model 14E1 \citep{Shigeyama_90} as shown in Figure
2. If the distribution of Ba is similar to that of $^{56}$Ni (i.e.,
the ratio Ba/$^{56}$Ni is uniform in the ejecta), the total amount of
Ba can be estimated from the mass fraction at the photosphere because
both the total mass and distribution of $^{56}$Ni are known. The
measurement of the mass fraction of Ba was performed based on the
spectrum obtained at day 20\footnote{Abundance analyses of Ba were
performed also for later spectra. However, as Mazzali \& Chugai (1995)
noted, the ionization by non-thermal electrons makes their model less
plausible in deeper layers of the SN ejecta that become visible at
later phases.}  and the mass fraction was found to be
$\sim2.4\times10^{-8}$\citep{Mazzali2_92,Mazzali_95} . According to
the above model 14E1, the photosphere at day 20 is located at the
enclosed mass of $\sim$12\ms (Fig. 2). Thus the mass of Ba is
estimated to be about $6\times10^{-6}$ \msp. A similar amount of Ba is
obtained based on another hydrodynamical model W10 \citep{Pinto88a,
Pinto88b, Pinto88c} that also derived the distribution of elements in
the ejecta from comparison with observations. The, Burrows, \& Bussard
(1990) also arrived at a similar distribution to obtain the light
curves consistent with the observed X-ray and $\gamma$-ray line
emissions at early phases based on hydrodynamical model 14E1.

\section{IMPLICATIONS FROM THE DERIVED BARIUM YIELD}

From two different angles, we could say that the inferred amount of Ba
$\sim 6\times10^{-6}$\ms in SN 1987A is significantly high. First of
all, this mass results in a large enhancement of Ba in comparison with
iron (Fe) in the ejecta. The mass of Fe contained in SN 1987A is known
with fairly good accuracy, because $^{56}$Ni eventually decays to
$^{56}$Fe.  As a result, the mass ratio of Ba to Fe in SN 1987A is found
to be a factor of $\sim 8$ larger than the corresponding solar abundance
ratio.

Recent studies on the chemical compositions of metal-poor halo stars
claim that these stars might inherit the abundance pattern of the ejecta
of the preceding few SNe \citep{Audouze_95} or a single SN
\citep{Shigeyama_98}. To put it another way, stars are likely to be
formed from the ISM comprising the ejecta of a single SN. Have we
already found the very star that has a large enhancement of Ba as
inferred in the ejecta of SN 1987A? The answer is yes -- The stars are
notable neutron-capture-rich giant stars CS22892-052 and CS31082-001,
whose abundance distributions of elements have been precisely determined
up to Z=90-92 \citep{Sneden_00,Hill_01}, and in which the
neutron-capture elements are found to be overabundant:
+0.3$<$[neutron-capture/Fe]$<$+1.8 ([X/Y] denotes the logarithmic value
of the number ratio of element X to element Y divided by the
corresponding solar ratio). The Ba/Fe ratios in these stars are a factor
of $\sim$8-12 larger than the corresponding solar ratio. This
coincidence of a large enhancement of Ba between SN 1987A and these two
stars might imply that these stars are descendants of SNe whose
progenitor masses were $\sim$20\msp. Here we compare the abundance
distribution of some elements in CS22892-052 with that in the ejecta of
SN 1987A (Fig.1). The masses of a few elements (Si, Ca, Ni) in SN 1987A
have been obtained from its nebular spectra \citep{Danziger_91}. For
deducing the mass of Sr, we use the same procedure as for Ba. The masses
thus obtained have been converted to the abundance ratios of element X
to Fe ([X/Fe]). We also show in this figure the abundance curve between
Mg and Ni predicted by recent nucleosynthesis calculations for 20\ms and
13\ms models \citep{Umeda_01}. The abundance distribution of CS22892-052
is very similar to that of SN 1987A, and also to the theoretical curve
of a 20 \ms nucleosynthesis model for lighter elements.

Secondly, the estimated mass of Ba implies that 20\ms SNe are the
predominant sites for $r$-process nucleosynthesis. This can be
illustrated by the following argument: Suppose that stars in the mass
range of $(20\pm\Delta M)$\ms yield the same amount of Ba as that from
SN 1987A, then what $\Delta M$ suffices to supply Ba to reach the
observed solar $r$-process abundance? Since current theoretical SN
models can predict the yields of oxygen as a function of the progenitor
mass with fairly good accuracy \citep{Tsujimoto_95,Woosley_95}, one can
estimate the mass ratio of these two elements ejected from all
SNe as Ba/O$\sim3\times10^{-7}\Delta M\big/M_\odot$, adopting the
Salpeter initial mass function. To explain the corresponding solar
abundance ratio by this SN yield, we deduce $\Delta
M$=0.7\msp. Therefore the number of stars populating the derived mass
range is only $\sim$4\% of all SN progenitor stars. This narrow
mass range suggests that the abundance distributions of $r$-process
elements in any star including the sun must be similar. This uniformity
in the abundance distributions of $r$-process elements has already been
pointed out by \citet{Cowan_99} in abundance analyses of four stars
CS22892-052, HD115444, HD122563, and HD126238 (in the latter three
stars, Ba is not so enhanced, i.e., [Ba/Fe]$<$+0.2), which led to the
conclusion that there is only one $r$-process site in the Galaxy, at
least for $Z\geq$56. Our finding is that the unique $r$-process site is
$(20\pm0.7)$\ms SNe, as represented by SN 1987A.

\section{CONNECTION BETWEEN SN 1987A AND NEUTRON-CAPTURE-RICH STARS}

A noteworthy point is that two stars $-$ CS22892-052 and CS31082-001 $-$
with a large enhancement of Ba such as [Ba/Fe]$\sim$+1.0 have similar
metallicities, i.e., [Fe/H]=$-3.1$, $-2.9$, respectively. This also
leads to the same conclusion as discussed above -- that these stars
inherit the abundance pattern of the 20 \ms SN belonging to the
first few generations in the Galaxy.  For a given SN in the early
galaxy, metallicities of the descendant stars can be estimated from the
ratio of the mass of a certain heavy element, conventionally Fe, ejected
from the SN, to the mass of hydrogen eventually swept up by the
explosion \citep{Shigeyama_98}. Neutron-capture-rich stars must be the
first few generation stars, because in later generation stars, such a
large enhancement of $r$-process elements would be reduced due to a
gradual increase in the contribution from the ISM with low
[$r$-process/Fe] ratios to the stellar abundance. Since each SN
with a different progenitor mass yields different amounts of heavy
elements, we obtain the stellar metallicity (or the Fe/H ratio) as a
function of the progenitor mass of a SN from which the star was
born. Taking a ratio of the Fe yield $\sim$0.07\ms to the swept-up mass
of hydrogen $\sim 6.5\times 10^4$ \ms for a 20 \ms SN, we obtain
[Fe/H]$\sim -3$. The metallicities of CS22892-052 and CS31082-001 indeed
suggest that their progenitor masses were $\sim$20 \msp.

Once the progenitor mass for these stars has been determined, we can
estimate the mass of Ba ejected by a 20 \ms SN, combining the
observed [Ba/Fe] ratios in these stars with the Fe yield from the
SN. In fact, it has been already found \citep{Tsujimoto2_00} that
CS22892-052 was born from a 20 \ms SN that had produced
$8\times10^{-6}$ \ms of Ba. Thus there is a fairly good agreement in the
Ba mass ejected from a 20 \ms SN independently predicted by both
SN 1987A and CS22892-052. It is a natural result of their similar
elemental abundance distributions.

Finally we need to examine Sr. The overabundance of Sr in comparison
with Fe in SN 1987A is relatively small, compared with the prediction by
theoretical $r$-process line \citep{Kappler_89} in Figure 1. This might
imply that Sr has other production sites besides 20\ms SNe.  This
consideration is compatible with the observed breakdown of the
concordance between the solar abundances and the CS22892-052 abundances
for the lighter $r$-process elements \citep{Sneden_00}. One possible
candidate for other sites may be a wider mass range for the production
site of Sr, e.g., $(20\pm3)$\msp. This range is obtained from the same
procedure for estimating the value of $\Delta M$ as was performed for
Ba. The obtained $\Delta M$ for Sr approximately corresponds to the
metallicity range $-3.3$\ltsim[Fe/H]\ltsim$-2.7$. It implies that among
stars in this metallicity range, there exist stars in which Sr is much
enhanced compared with Ba. As a matter of fact, HD122563 ([Fe/H]=--2.74)
has a very large enhancement of Sr such as [Sr/Ba]=+1.1
\citep{Westin_00}.

\section{CONCLUSIONS}

We have shown that $r$-process nucleosynthesis is certainly evident in
SN 1987A. Our study clarifies that the observed overabundance of Ba in
the outer layer of the ejecta of SN 1987A is part of the $r$-process
production synthesized during explosion in the deepest layers of the
ejecta, though this overabundance was ascribed to the result of the
$s$-process operating in the core He burning stage. We presented three
strong arguments for this.

First, we should pay attention to the observed fact that the
overabundance of Sr was also found at the same time in SN 1987A and
that the abundance of Sr is less enhanced than that of Ba, which makes
it unlikely that these elements are of $s$-process origin. In
addition, the derived Ba/Sr ratio in SN 1987A is rather similar to
those observed from metal-poor stars in which Ba has been shown to be
of $r$-process origin from the observed Ba/Eu ratios.

Secondly, the absence of an overabundance of Ba in other SNe II-P
contradicts the prediction from theoretical models that the $s$-process
operates {\it irrespective of} He-core masses. Alternatively, this fact
implies that the $r$-process occurs in SNe whose progenitor
masses are limited to a narrow range. Such a unique production site for
$r$-process is fully consistent with the uniformity in the abundance
distributions of $r$-process elements among several stars including the
sun.

Finally, the total amount of Ba inferred from the Ba abundance in the
outer layer of the ejecta from the standpoint of the $r$-process origin
reveals that 20 \ms SNe, one of which is SN 1987A, are the
prominent production sites for $r$-process elements. This is consistent
with the second point above, i.e., no overabundances of Ba in type II-P
SNe other than SN 1987A and the uniformity in the stellar
abundance distributions of $r$-process elements.

We have further shown that the elemental abundance distributions of
neutron-capture-rich giant stars $-$ CS22892-052 and CS31082-001 $-$ are
very similar to that of SN 1987A, which suggests that these stars are
born from the ejecta of 20 \ms SNe like SN 1987A in the early
Galaxy. The high abundances of Ba observed in these stars also imply
that the major production sites for $r$-process are $\sim$20 \ms
SNe.

\bigskip

This work has been partly supported by COE research (07CE2002) and a
Grant-in-Aid for Scientific Research (12640242) of the Ministry of
Education, Science, Culture, and Sports in Japan.

\end{document}